\documentclass{ws-ijmpcs}

\begin{document}

\markboth{Yu.S.~Surovtsev, P.~Byd\v{z}ovsk\'y, T.~Gutsche, R.~Kami\'nski, V.E.~Lyubovitskij, M.~Nagy}
{Effect of the multi-channel $\pi\pi$ scattering in two-pion transitions of the $\Upsilon$'s }

%
\catchline{}{}{}{}{}
%

\title{Effect of coupled channels of the multi-channel pion-pion scattering in two-pion transitions of the $\Upsilon$ mesons}

\author{Yurii S.~Surovtsev}

\address{\it Bogoliubov Laboratory of Theoretical Physics, Joint Institute for Nuclear Research, 141980 Dubna, Russia\\
surovcev@theor.jinr.ru}

\author{Petr Byd\v{z}ovsk\'y}

\address{Nuclear Physics Institute, Academy of Sciences of the Czech Republic, 250 68 \v{R}e\v{z} near Prague, Czech Republic\\
bydz@ujf.cas.cz}

\author{Thomas Gutsche}

\address{Institut f\"ur Theoretische Physik, Universit\"at T\"ubingen, Kepler Center for Astro and Particle Physics,
Auf der Morgenstelle 14, D-72076 T\"ubingen, Germany\\
gutsche@uni-tuebingen.de}

\author{Robert~Kami\'nski}

\address{Institute of Nuclear Physics of the Polish Academy of Sciences, Cracow 31342, Poland\\
Robert.Kaminski@ifj.edu.pl}

\author{Valery E. Lyubovitskij}

\address{Institut f\"ur Theoretische Physik, Universit\"at T\"ubingen, Kepler Center for Astro and Particle Physics,
Auf der Morgenstelle 14, D-72076 T\"ubingen, Germany\\
Department of Physics, Tomsk State University, 634050 Tomsk, Russia\\
Mathematical Physics Department, Tomsk Polytechnic University, Lenin Avenue 30, 634050 Tomsk, Russia\\
lubovit@tphys.physik.uni-tuebingen.de}

\author{Miroslav Nagy}

\address{Institute of Physics, Slovak Academy of Sciences, D\'ubravsk\'a cesta 9, Bratislava 84511, Slovak Republic\\
miroslav.nagy@savba.sk}

\maketitle

\begin{history}
\received{Day Month Year}
\revised{Day Month Year}
\published{Day Month Year}
\end{history}
\newpage
\begin{abstract}
The effect of isoscalar S-wave processes $\pi\pi\to\pi\pi,K\overline{K},\eta\eta$ is considered in the analysis of data (from the ARGUS, CLEO, CUSB, Crystal Ball, Belle, {\it BABAR} collaborations) on the bottomonia decays --- $\Upsilon(mS)\to\Upsilon(nS)\pi\pi$ ($m>n, m=2,3,4,5, n=1,2,3$). It is shown that the dipion mass spectra of these decays are explained by the unified mechanism related to the contribution of the multichannel $\pi\pi$ scattering in the final states of these reactions. Since in the analysis the multichannel $\pi\pi$-scattering amplitudes did not change in comparison with the ones in our combined analysis of data on the multichannel $\pi\pi$ scattering and on the charmonia decays  --- $J/\psi\to\phi(\pi\pi, K\overline{K})$ and $\psi(2S)\to J/\psi(\pi\pi)$ (from the Crystal Ball, DM2, Mark~II, Mark~III, and BES~II) --- the results confirm all our earlier conclusions on the scalar mesons.

\end{abstract}

\ccode{PACS numbers: 11.55.Bq,11.80.Gw,12.39.Mk,14.40.Pq}

\section{Introduction}	
In the analysis of practically all available data on two-pion transitions of the $\Upsilon$ mesons from the ARGUS, CLEO, CUSB, Crystal Ball, Belle, and {\it BABAR} collaborations ---  $\Upsilon(mS)\to\Upsilon(nS)\pi\pi$ ($m>n$, $m=2,3,4,5,$ $n=1,2,3$) --- the contribution of multi-channel $\pi\pi$ scattering in the final-state interactions is considered.
The analysis, aimed at studying the scalar mesons, is performed jointly considering the isoscalar S-wave processes $\pi\pi\to\pi\pi,K\overline{K},\eta\eta$, which are described in our approach based on analyticity and unitarity and using an uniformization procedure, and the charmonia decays --- $J/\psi\to\phi(\pi\pi, K\overline{K})$, $\psi(2S)\to J/\psi\pi\pi$ --- from the Crystal Ball, DM2, Mark~II, Mark~III, and BES~II collaborations.
Possibility of using two-pion transitions of heavy quarkonia as a good laboratory for studying the $f_0$ mesons is related to the expected fact that the dipion is produced in a $S$-wave whereas the final quarkonium is a spectator\cite{MP-prd93}. The problem of interpretation of the scalar mesons is faraway to be solved completely\cite{PDG-14}.

On the other hand, an explanation of the dipion mass distributions of the $\Upsilon(mS)$ where $m>2$ contains a number of surprises,
discussion of which for the $\Upsilon(3S)$ decays and our explanation are contained in Ref.~\refcite{SBGKLN-prd15}.
In one's turn, the $\Upsilon(4S)$ and $\Upsilon(5S)$ are distinguished from the lower $\Upsilon$-states by the fact that their masses are above the $B\overline{B}$ threshold. The dipion mass distributions of these decays have the additional mysteries, e.g. the sharp dips about 1~GeV in the two-pion transitions of these states to the basic ones.
These higher states predominantly decay into pairs of the $B$-meson family because these modes are not suppressed by the OZI rule: the $\Upsilon(4S)$ decays into $B\overline{B}$ pairs form more than $96$\% of the total width; for the $\Upsilon(5S)$ these decay modes make up about $90$\%.
The rates of decay modes of interest are $(8.1\pm 0.6)\times 10^{-3}$\% and $(8.6\pm1.3)\times 10^{-3}$\% of the total width for $\Upsilon(4S)\to\Upsilon(1S)\pi\pi$ and $\Upsilon(4S)\to\Upsilon(2S)\pi\pi$, respectively, and about $(0.5\div0.8)$\% for $\Upsilon(5S)\to\Upsilon(1S,2S,3S)\pi\pi$ (see Ref.~\refcite{PDG-14}). The total widths of $\Upsilon(5S)$ and $\Upsilon(4S)$ are 110 and 20.5~MeV, respectively, and the one of the $\Upsilon(3S)$ is only 20.32~keV. The partial decay widths of $\Upsilon(4S)\to\Upsilon(1S,2S)\pi\pi$ are almost of the same order as the ones of the decays $\Upsilon(3S)\to\Upsilon(1S,2S)\pi\pi$ which form about $(2\div4.5)$\% of the total widths.  The widths of decays $\Upsilon(5S)\to\Upsilon(1S,2S,3S)\pi\pi$ are larger than the latter ones by about $2\div3$ orders of magnitude. These facts might point to interesting dynamics which is responsible for the coupling of the decay $\Upsilon$ mesons with the final-state particles and cannot be related only to the $B\bar B$ transition dynamics.

We shall show that in the two-pion transitions of $\Upsilon(mS)$ ($m=2,3,4,5$) the basic mechanism which explains the characteristic shapes of dipion mass distributions is unified for the two-pion transitions both of bottomonia and charmonia,
i.e., it is not related to the $B\bar B$ transition dynamics. It is based on our previous conclusions on the wide resonances\cite{SBLKN-prd14,SBKLN-PRD12} and is related to the interference of the contributions of multichannel $\pi\pi$ scattering in the final state.

We also work out the role of the individual $f_0$ resonances in contributing to the dipion mass distributions in the decays $\Upsilon(mS)\to\Upsilon(nS) \pi^+ \pi^-$ ($m>n$, $m=3,4,5;~n=1,2,3$). For this purpose we first discuss some results from our previous paper\cite{SBLKN-prd14}.

\section{The model-independent amplitudes for multichannel $\pi\pi$ scattering}

As multichannel $\pi\pi$ scattering there are considered reactions $\pi\pi\to\pi\pi,K\overline{K},\eta\eta$, i.e. the three coupled channels. The 3-channel $S$-matrix is determined on the 8-sheeted Riemann surface. The matrix elements $S_{ij}$, where $i,j=1,2,3$ denote the channels, have right-hand cuts along the real axis of the complex $s$ plane ($s$ is the invariant total energy squared), starting with the channel thresholds $s_i$ ($i=1,2,3$), and the left-hand cuts.
The Riemann surface structure can be represented by taking the following uniformizing variable\cite{SBL-prd12}:
\begin{equation}
w=\bigl[\sqrt{(s-s_2)s_3} + \sqrt{(s-s_3)s_2}\bigl]/\sqrt{s(s_3-s_2)}\,
\end{equation}
with $s_2=4m_K^2$ and $s_3=4m_\eta^2$, where we have neglected the $\pi\pi$-threshold branch point and included the $K\overline{K}$- and $\eta\eta$-threshold branch-points and the left-hand branch-point at $s=0$ related to the crossed channels.

Resonance representations on the Riemann surface are obtained using formulas from Ref.~\refcite{SBL-prd12}. Analytic continuations of the $S$-matrix elements to all sheets are expressed in terms of those on the physical (I) sheet that have only the resonance zeros (beyond the real axis). The multichannel resonances are classified according to these resonance zeros on sheet~I. In the 3-channel case there are {\it seven types} of resonances corresponding to seven possible situations when there are resonance zeros on sheet I only in $S_{11}$ -- ({\bf a}); ~~$S_{22}$ -- ({\bf b}); ~~$S_{33}$ -- ({\bf c}); ~~$S_{11}$ and $S_{22}$ -- ({\bf d}); ~~$S_{22}$ and $S_{33}$ -- ({\bf e}); ~~$S_{11}$ and $S_{33}$ -- ({\bf f}); ~~$S_{11}$, $S_{22}$ and $S_{33}$ -- ({\bf g}). The resonance of every type is represented by a pair of complex-conjugate {\it clusters} (of poles and zeros on the Riemann surface).
The $S$-matrix elements $S_{ij}$ are parameterized using the Le~Couteur--Newton relations\cite{LeCou}. They express the $S$-matrix elements of all coupled processes in terms of the Jost determinant $d(\sqrt{s-s_1},\cdots,\sqrt{s-s_n})$ which
is a real analytic function with the only branch points at $\sqrt{s-s_i}=0$. The $S$-matrix elements are taken as the products $S=S_B S_{res}$; the main ({\it model-independent}) contribution of resonances, given by the pole clusters, is included in the resonance part $S_{res}$; possible remaining small ({\it model-dependent}) contributions of resonances and the influence of channels which are not taken explicitly into account in the uniformizing variable are included in the background part $S_B$. The $d_{res}(w)$-function for the resonance part, which now is free from any branch points, is taken as
\begin{equation}
d_{res}(w)=w^{-\frac{M}{2}}\prod_{r=1}^{M}(w+w_{r}^*)
\end{equation}
where $M$ is the number of resonance zeros. For the background part we have
\begin{equation}
d_B=\mbox{exp}[-i\sum_{n=1}^{3}(\sqrt{s-s_n}/2m_n)(\alpha_n+i\beta_n)]
\end{equation}
with
\begin{eqnarray}	
\begin{array}{rcl}
&&\alpha_n=a_{n1}+a_{n\sigma}\frac{s-s_\sigma}{s_\sigma}\theta(s-s_\sigma)+
a_{nv}\frac{s-s_v}{s_v}\theta(s-s_v),\\[8pt]
&&\beta_n=b_{n1}+b_{n\sigma}\frac{s-s_\sigma}{s_\sigma}\theta(s-s_\sigma)+
b_{nv}\frac{s-s_v}{s_v}\theta(s-s_v)
\end{array}
\end{eqnarray}
where $s_\sigma$ is the $\sigma\sigma$ threshold, $s_v$ the combined threshold of the $\eta\eta^{\prime},~\rho\rho,~\omega\omega$ channels.
The resonance zeros $w_{r}$ and the background parameters were fixed by fitting to the data on $\pi\pi\to\pi\pi,K\overline{K},\eta\eta$ and the charmonium decay processes --- $J/\psi\to\phi(\pi\pi, K\overline{K})$, $\psi(2S)\to J/\psi\pi\pi$\cite{SBLKN-prd14}.

The preferred scenario found is when the $f_0(500)$ is described by the cluster of type ({\bf a}); the $f_0(1370)$, $f_0(1500)$ and $f_0(1710)$ with type ({\bf c}); and $f_0^\prime(1500)$ by type ({\bf g}); the $f_0(980)$ is represented only by the pole on sheet~II and a shifted pole on sheet~III. The obtained pole-clusters for the resonances can be found in Ref.~\refcite{SBLKN-prd14}.
The obtained background parameters are: $a_{11}=0.0$, $a_{1\sigma}=0.0199$, $a_{1v}=0.0$, $b_{11}=b_{1\sigma}=0.0$,
$b_{1v}=0.0338$; $a_{21}=-2.4649$, $a_{2\sigma}=-2.3222$, $a_{2v}=-6.611$, $b_{21}=b_{2\sigma}=0.0$, $b_{2v}=7.073$; $b_{31}=0.6421$, $b_{3\sigma}=0.4851$; $b_{3v}=0$; $s_\sigma=1.6338~{\rm GeV}^2$, $s_v=2.0857~{\rm GeV}^2$.

Let us indicate some important conclusions from the fact of obtaining the small (zero for the elastic region) values of the $\pi\pi$-scattering background parameters derived after allowing for the left-hand branch-point at $s=0$:
1) The confirmation of our assumption $S=S_B S_{res}$.
2) The indication that the representation of multi-channel resonances by the pole clusters on the uniformization plane is good and quite sufficient.
3) The indication that the consideration of the left-hand branch-point at $s=0$ in the uniformizing variable partly solves a problem of some approaches (see, e.g., Ref.~\refcite{Achasov-Shest}) where the wide-resonance parameters are strongly controlled by the non-resonant background.
4) Since the contribution to the $\pi\pi$ scattering amplitude from the crossed channels is given by allowing for the left-hand branch-point at $s=0$ in the uniformizing variable and the meson-exchange contributions in the left-hand cuts, the zero background in the elastic-scattering region obtained in the 3-channel analysis of the processes $\pi\pi\to\pi\pi,K\overline{K},\eta\eta$ indicates that the $\rho$- and $f_0(500)$-meson exchange contributions in the left-hand cut practically cancel each other. One can show allowing for gauge invariance that the vector- and scalar-meson exchanges contribute with opposite signs. Therefore, the practically zero background in $\pi\pi$ scattering is an additional confirmation that the $f_0(500)$, observed in the analysis as the pole cluster of type {\bf a}, is indeed a particle (though very wide), not some dynamically formed resonance.
5) Finally, a reasonable and simple description of the background should be a criterion for the correctness of the approach.

Generally, {\it wide multi-channel states are most adequately represented by pole clusters}, because the pole clusters give the main model-independent effect of resonances. As to such parameters of the wide multi-channel states, as masses, total widths and coupling constants with channels, they should be calculated using the poles on sheets II, IV and VIII, because only on these sheets the analytic continuations have the forms: $\propto 1/S_{11}^{\rm I}$, $\propto 1/S_{22}^{\rm I}$ and $\propto 1/S_{33}^{\rm I}$, respectively, i.e., the pole positions of resonances are at the same points of the complex-energy plane, as the resonance zeros on the physical sheet, and are not shifted due to the coupling of channels. E.g., if the resonance part of amplitude is taken as
~$T^{res}=\sqrt{s}~\Gamma_{el}/(m_{res}^2-s-i\sqrt{s}~\Gamma_{tot})$, for the mass and total width, one obtains
$m_{res}=\sqrt{{\rm E}_r^2+\left(\Gamma_r/2\right)^2}~{\rm and}~~~\Gamma_{tot}=\Gamma_r$,
where the pole position $\sqrt{s_r}\!=\!{\rm E}_r\!-\!i\Gamma_r/2$ must be taken on sheets II, IV, VIII, depending on the resonance classification. In Table~\ref{tab:mass} we show the obtained masses and total widths of the $f_0$ resonances\cite{SBLKN-prd14}.
\begin{table}[h]
\tbl{The masses and total widths of the $f_0$ resonances.}
{\begin{tabular}{@{}ccccccc@{}} \toprule {} & $f_0(500)$ & $f_0(980)$ & $f_0(1370)$ & $f_0(1500)$ & $f_0^\prime(1500)$ & $f_0(1710)$\\ \colrule
$m_{res}$[MeV] & 693.9$\pm$10.0 & 1008.1$\pm$3.1 & 1399.0$\pm$24.7 & 1495.2$\pm$3.2 & 1539.5$\pm$5.4 & 1733.8$\pm$43.2 \\ \colrule
$\Gamma_{tot}$[MeV] & 931.2$\pm$11.8 & 64.0$\pm$3.0
& 357.0$\pm$74.4 & 124.4$\pm$18.4 & 571.6$\pm$25.8 & 117.6$\pm$32.8 \\
\botrule
\end{tabular} \label{tab:mass}}
\end{table}

We further investigated the role of the individual $f_0$ resonances in contributing to the shape of the di-meson mass distributions in the bottomonia decays. It is reasonable to switch off only those resonances [$f_0(500)$, $f_0(1370)$, $f_0(1500)$ and $f_0(1710)$], removal of which can be somehow compensated by correcting the background to have the more-or-less acceptable description of the multi-channel $\pi\pi$ scattering.
First, when leaving out before-mentioned resonances, a minimal set of the $f_0$ mesons consisting of the $f_0(500)$, $f_0(980)$, and $f_0^\prime(1500)$ is sufficient to achieve a description of the processes
$\pi\pi\to\pi\pi,K\overline{K},\eta\eta$ with a total $\chi^2/\mbox{ndf}\approx1.20$. The obtained, adjusted background parameters are: $a_{11}=0.0$, $a_{1\sigma}=0.0321$, $a_{1v}=0.0$, $b_{11}=-0.0051$, $b_{1\sigma}=0.0$, $b_{1v}=0.04$; $a_{21}=-1.6425$, $a_{2\sigma}=-0.3907$, $a_{2v}=-7.274$, $b_{21}=0.1189$, $b_{2\sigma}=0.2741$, $b_{2v}=5.823$; $b_{31}=0.7711$, $b_{3\sigma}=0.505$, $b_{3v}=0.0$.
Second, from these three mesons only the $f_0(500)$ can be switched off while still obtaining a reasonable description of multichannel $\pi\pi$ scattering (though with an appearance of the pseudo-background) with a total $\chi^2/\mbox{ndf}\approx1.43$ and with the following corrected background parameters: $a_{11}=0.3513$, $a_{1\sigma}=-0.2055$, $a_{1v}=0.207$, $b_{11}=-0.0077$, $b_{1\sigma}=0.0$, $b_{1v}=0.0378$; $a_{21}=-1.8597$, $a_{2\sigma}=0.1688$, $a_{2v}=-7.519$, $b_{21}=0.161$, $b_{2\sigma}=0.0$, $b_{2v}=6.94$; $b_{31}=0.7758$, $b_{3\sigma}=0.4985$, $b_{3v}=0.0$.

\section{The contribution of multi-channel $\pi\pi$ scattering in the final states of decays of $\Upsilon$-meson families}

For decays $J/\psi\to\phi\pi\pi,\phi K\overline{K}$ we have taken data from Mark~III, from DM2 and from BES~II; for $\psi(2S)\to J/\psi(\pi^+\pi^-)$ from Mark~II; for $\psi(2S)\to J/\psi(\pi^0\pi^0)$ from Crystal Ball(80) collaborations (the references to the used data on the charmonia decays are in Ref.~\refcite{SBLKN-prd14}).
For $\Upsilon(2S)\to\Upsilon(1S)(\pi^+\pi^-,\pi^0\pi^0)$ data were taken from ARGUS\cite{Argus}, CLEO\cite{CLEO,CLEO07}, CUSB\cite{CUSB}, and the Crystal Ball\cite{Crystal_Ball(85)} collaborations.
For $\Upsilon(3S)\to\Upsilon(1S)(\pi^+\pi^-,\pi^0\pi^0)$ and $\Upsilon(3S)\to\Upsilon(2S)(\pi^+\pi^-,\pi^0\pi^0)$ measurements are available from the CLEO collaboration\cite{CLEO07,CLEO(94)}; finally, for the decays
$\Upsilon(4S,5S)\to\Upsilon(ns)\pi^+ \pi^-$ ($n=1,2,3$) from the {\it BABAR}\cite{BaBar06} and Belle\cite{Belle} collaborations.

The used formalism for calculating the di-meson mass distributions in the $\Upsilon(mS)$ decays is analogous to the one proposed in Ref.~\refcite{MP-prd93} for the decays $J/\psi\to\phi(\pi\pi, K\overline{K})$ and $V^{\prime}\to V\pi\pi$ ($V=\psi,\Upsilon$).
I.e., it was assumed that the pion pairs in the final state have zero isospin and spin. Only these pairs of pions undergo final state interactions whereas the final $\Upsilon(nS)$ meson ($n<m$) acts as a spectator. The amplitudes for the decays $\Upsilon(mS)\to\Upsilon(nS)\pi\pi$ ($m>n$, $m=2,3,4,5,$ $n=1,2,3$) include the scattering amplitudes $T_{ij}$
$(i,j=1-\pi\pi,2-K\overline{K})$ as follows
\begin{equation}
F_{mn}(s) = (\rho_{mn}^0+\rho_{mn}^1\,s)\,T_{11}
+ (\omega_{mn}^0+\omega_{mn}^1\,s)\,T_{21},
\end{equation}
where indices $m$ and $n$ correspond to $\Upsilon(mS)$ and $\Upsilon(nS)$, respectively.
The free parameters $\rho_{mn}^0$, $\rho_{mn}^1$, $\omega_{mn}^0$, and $\omega_{mn}^1$
depend on the couplings of the $\Upsilon(mS)$ to the channels $\pi\pi$ and $K\overline{K}$. The model-independent amplitudes $T_{ij}$ are expressed through the $S$-matrix elements shown in the previous section as
\begin{equation}
S_{ij}=\delta_{ij}+2i\sqrt{\rho_1\rho_2}T_{ij}
\end{equation}
where $\rho_i=\sqrt{1-s_i/s}$ and $s_i$ is the reaction threshold. The expressions for the dipion mass distributions in the decay $\Upsilon(mS)\to\Upsilon(nS)\pi\pi$ are
\begin{equation}
N|F|^{2}\sqrt{(s-s_1) \lambda(m_{\Upsilon(mS)}^2,s,m_{\Upsilon(nS)}^2)}\,,
\end{equation}
where $\lambda(x,y,z)=x^2+y^2+z^2-2xy-2yz-2xz$ is the K\"allen function.
The normalization $N$ is determined by a fit to the specific experiment:
for $\Upsilon(2S)\to \Upsilon(1S)\pi^+\pi^-$,
4.3439 for ARGUS\cite{Argus}, 2.1776 for CLEO(84)\cite{CLEO}, 1.2011 for CUSB\cite{CUSB};
for $\Upsilon(2S)\to\Upsilon(1S)\pi^0\pi^0$, 0.0788 for Crystal Ball(85)\cite{Crystal_Ball(85)};
for $\Upsilon(3S)\to\Upsilon(1S)(\pi^+\pi^-~{\rm and}~\pi^0\pi^0)$,
0.5096 and 0.2235 for CLEO(07)\cite{CLEO07}, and for $\Upsilon(3S)\to\Upsilon(2S)(\pi^+\pi^-$ ${\rm and}~\pi^0\pi^0)$,
2.0302 and 1.0316 for CLEO(94)\cite{CLEO(94)}, respectively;
for $\Upsilon(4S)\to\Upsilon(1S)\pi\pi$, ~7.1476 for {\it BABAR}(06)\cite{BaBar06} and ~0.5553 for Belle(07)\cite{Belle};
for $\Upsilon(4S)\to\Upsilon(2S)\pi\pi$, ~58.143 for {\it BABAR}(06);
for $\Upsilon(5S)\to\Upsilon(1S)\pi\pi$, $\Upsilon(5S)\to\Upsilon(2S)\pi\pi$
and $\Upsilon(5S)\to\Upsilon(3S)\pi\pi$, respectively, ~0.1626, 4.8355
and 10.858 for Belle(12)\cite{Belle}.

The~obtained~parameters~of~the~coupling~functions~of~the~decay~particles~$\Upsilon(mS)~(m=2,...,5)$
to channel~$i$ are:
\begin{eqnarray}
&&(\rho^{0}_{21},\rho^{1}_{21},\omega^{0}_{21},\omega^{1}_{21})=
(0.4050, 47.0963, 1.3352,-21.4343),\nonumber\\ &&(\rho^{0}_{31},\rho^{1}_{31},\omega^{0}_{31},\omega^{1}_{31})=
(1.0827,-2.7546,~ 0.8615, ~0.6600),\nonumber\\ &&(\rho^{0}_{32},\rho^{1}_{32},\omega^{0}_{32},\omega^{1}_{32})=
(1.6409,~1.8581,-1.2533,-40.0571),\nonumber\\ &&(\rho^{0}_{41},\rho^{1}_{41},\omega^{0}_{41},\omega^{1}_{41})=
(0.6162,-2.5715,-0.8467,~0.2128),\nonumber\\ &&(\rho^{0}_{42},\rho^{1}_{42},\omega^{0}_{42},\omega^{1}_{42})=
(2.329,-7.3511,~1.8096,-10.1477),\nonumber\\ &&(\rho^{0}_{51},\rho^{1}_{51},\omega^{0}_{51},\omega^{1}_{51})=
(0.7078,~4.0132,~4.838,-3.9091),\nonumber\\ &&(\rho^{0}_{52},\rho^{1}_{52},\omega^{0}_{52},\omega^{1}_{52})=
(0.8133,~2.2061,-0.7973,~0.3247),\nonumber\\ &&(\rho^{0}_{53},\rho^{1}_{53},\omega^{0}_{53},\omega^{1}_{53})=
(0.8946,~2.538,~0.627,-0.0483).\nonumber
\end{eqnarray}
The parameters for the decays of charmonia in this combined analysis did not change in comparison with our analysis without the data on bottomonia decays and can be found in Ref.~\refcite{SBLKN-prd14}.

A satisfactory combined description of all considered processes is obtained
with a total $\chi^2/\mbox{ndf}=804.677/(714-94)\approx1.30$;
for $\pi\pi$ scattering, $\chi^2/\mbox{ndf}\approx1.15$;
for $\pi\pi\to K\overline{K}$, $\chi^2/\mbox{ndf}\approx1.65$;
for $\pi\pi\to\eta\eta$, $\chi^2/\mbox{ndp}\approx0.87$;
for the decays $J/\psi\to\phi(\pi^+\pi^-, K^+K^-)$, $\chi^2/\mbox{ndp}\approx1.36$;
for $\psi(2S)\to J/\psi(\pi^+\pi^-,\pi^0\pi^0)$, $\chi^2/\mbox{ndp}\approx2.43$;
for $\Upsilon(2S)\to\Upsilon(1S)(\pi^+\pi^-,\pi^0\pi^0)$, $\chi^2/\mbox{ndp}\approx1.01$;
for $\Upsilon(3S)\to\Upsilon(1S)(\pi^+\pi^-,\pi^0\pi^0)$, $\chi^2/\mbox{ndp}\approx0.67$,
for $\Upsilon(3S)\to\Upsilon(2S)(\pi^+\pi^-,\pi^0\pi^0)$, $\chi^2/\mbox{ndp}\approx0.50$,
for $\Upsilon(4S)\to\Upsilon(1S)(\pi^+\pi^-)$, $\chi^2/\mbox{ndp}\approx0.27$,
for $\Upsilon(4S)\to\Upsilon(2S)(\pi^+\pi^-)$, $\chi^2/\mbox{ndp}\approx0.27$,
for $\Upsilon(5S)\to\Upsilon(1S)(\pi^+\pi^-)$, $\chi^2/\mbox{ndp}\approx1.80$,
for $\Upsilon(5S)\to\Upsilon(2S)(\pi^+\pi^-)$, $\chi^2/\mbox{ndp}\approx1.08$,
for $\Upsilon(5S)\to\Upsilon(3S)(\pi^+\pi^-)$, $\chi^2/\mbox{ndp}\approx0.81$.
Here "ndp" is the number of data points.

\begin{figure}[p]
\begin{center}
\includegraphics[width=0.44\textwidth,angle=0]{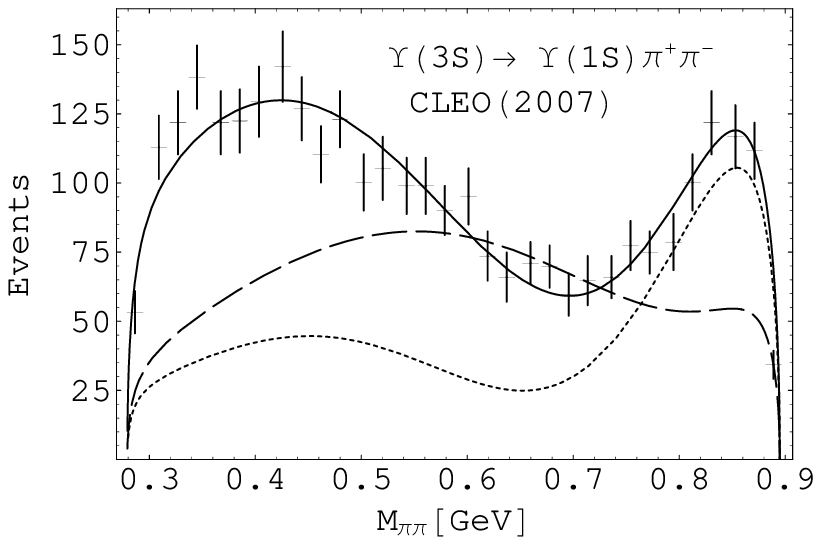}
\includegraphics[width=0.44\textwidth,angle=0]{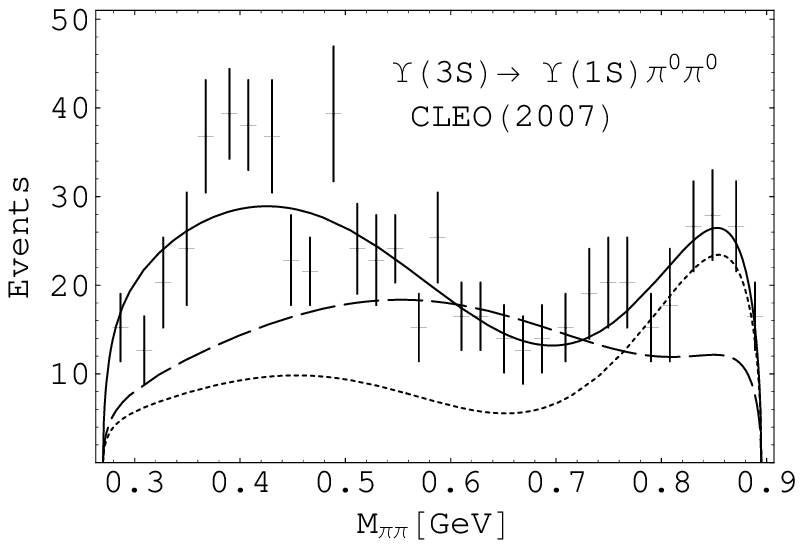}\\
\vspace*{0.1cm}
\includegraphics[width=0.44\textwidth,angle=0]{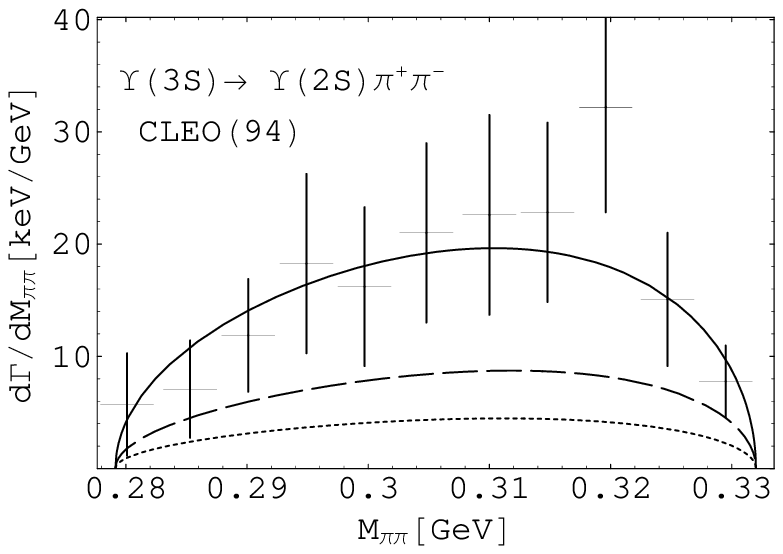}
\includegraphics[width=0.44\textwidth,angle=0]{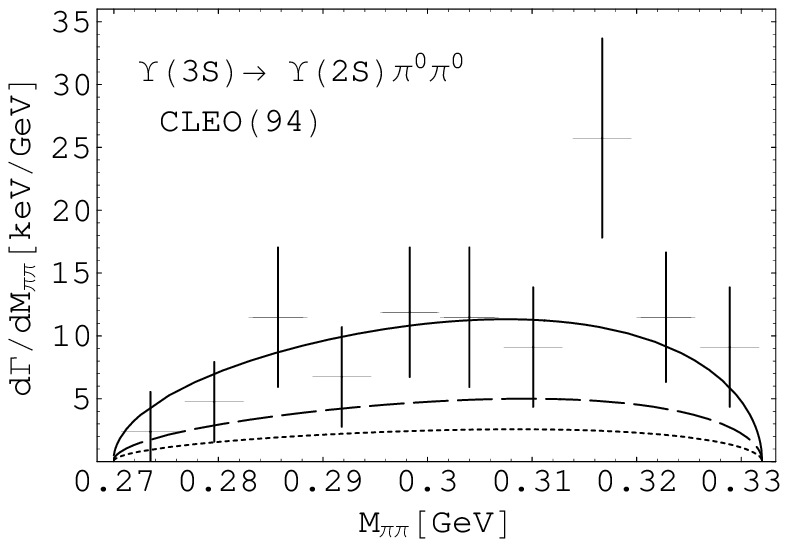}\\
\vspace*{0.1cm}
\includegraphics[width=0.44\textwidth,angle=0]{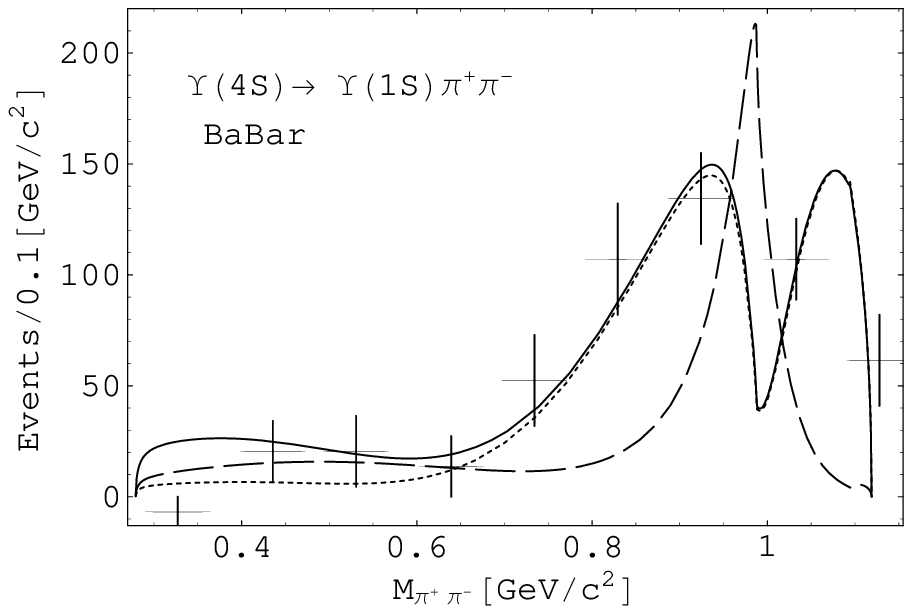}
\includegraphics[width=0.44\textwidth,angle=0]{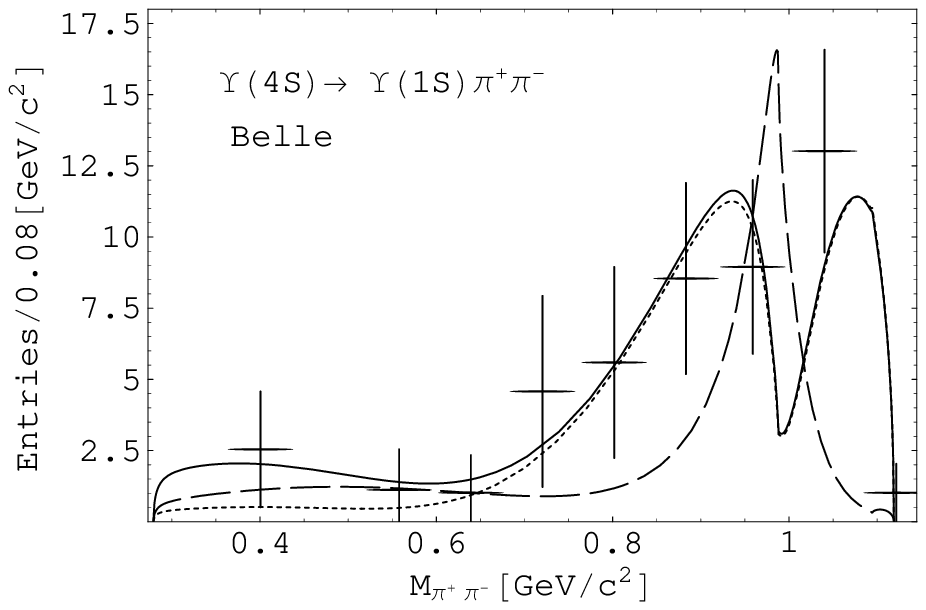}\\
\vspace*{0.1cm}
\includegraphics[width=0.44\textwidth,angle=0]{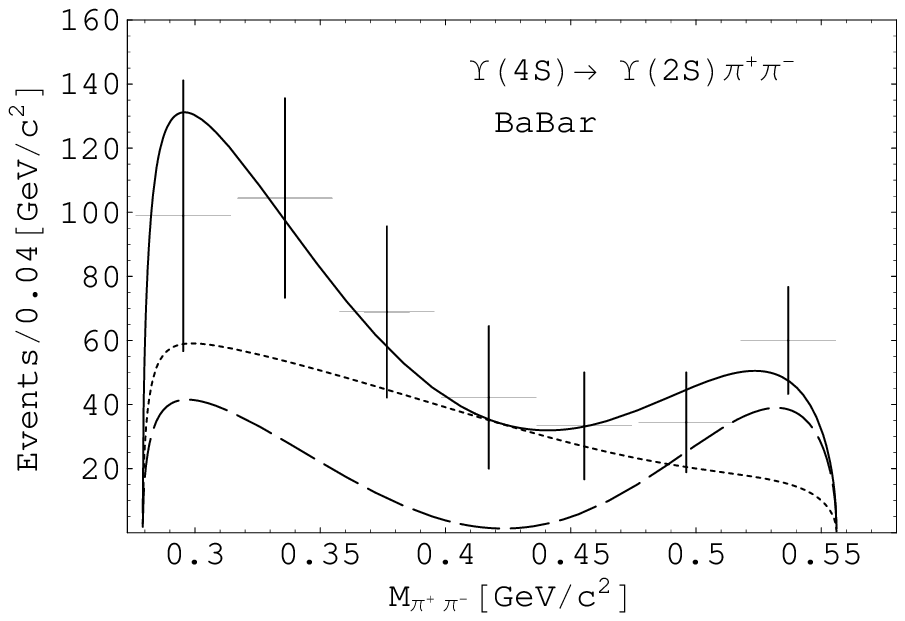}
\includegraphics[width=0.44\textwidth,angle=0]{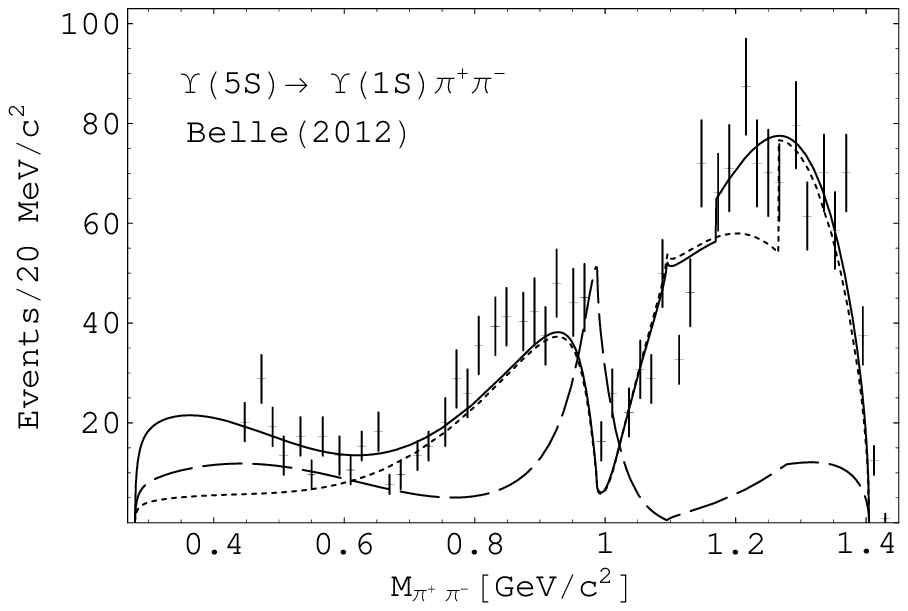}\\
\vspace*{0.1cm}
\includegraphics[width=0.44\textwidth,angle=0]{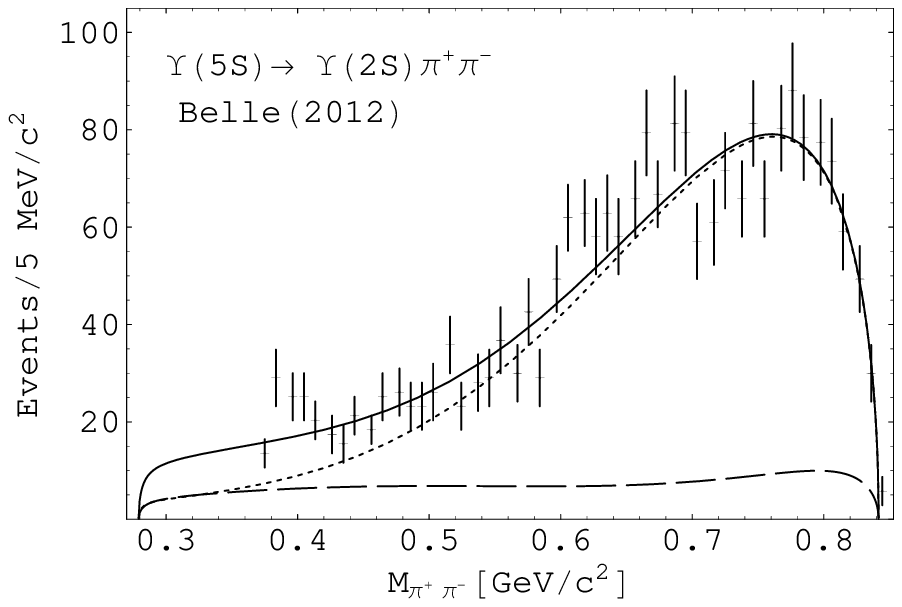}
\includegraphics[width=0.44\textwidth,angle=0]{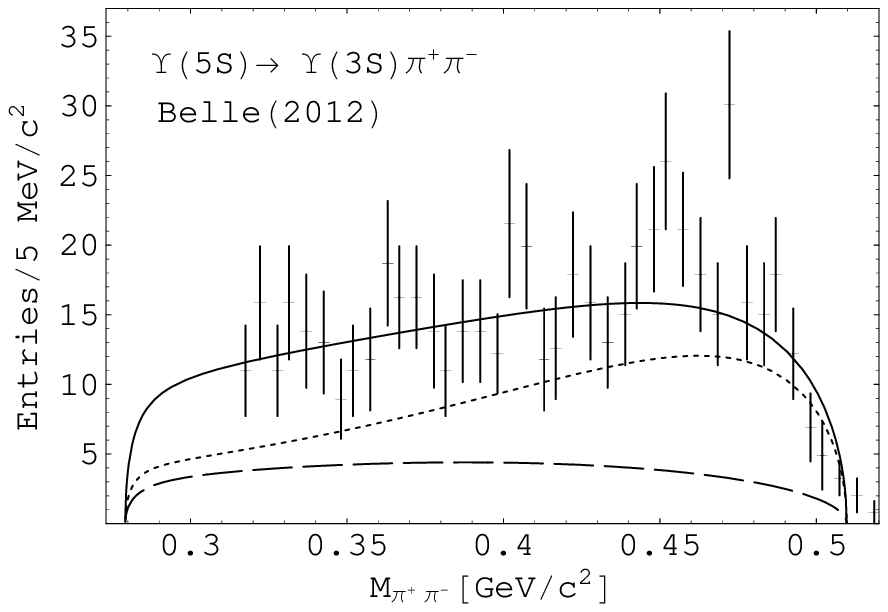}
\vspace*{4pt}
\caption{The decays $\Upsilon(mS)\to\Upsilon(nS)\pi\pi$ ($m>n$, $m=3,4,5,$ $n=1,2,3$). The solid lines correspond to the contribution of all relevant resonances; the dotted, of the $f_0(500)$, $f_0(980)$, and $f_0^\prime(1500)$; the dashed, of the $f_0(980)$ and $f_0^\prime(1500)$. \label{fig:ups_mn}}
\end{center}
\end{figure}

In Fig.~\ref{fig:ups_mn} we show the fits (solid lines) to the experimental data on the decays of bottomonia --- $\Upsilon(mS)\to\Upsilon(nS)\pi\pi$ ($m>n$,~$m=3,4,5,$~$n=1,2,3$). In these figures the dotted and dashed lines correspond to allowing for the contributions of the $f_0(500)$, $f_0(980)$, and $f_0^\prime(1500)$ and of the $f_0(980)$ and $f_0^\prime(1500)$, respectively, and the other $f_0$ resonances switched off.
The combined analysis including decay data on $J/\psi\to\phi(\pi\pi,K\overline{K})$
and $\psi(2S)\to J/\psi\pi\pi$ from the Mark~III, DM2, BES~II, Mark~II and Crystal Ball collaborations (see corresponding references also in Ref.~\refcite{SBLKN-prd14}) was found to be important for getting unique solutions to the $f_0$-meson parameters: first, we solved the ambiguity in the parameters of the $f_0(500)$\cite{SBL-prd12} in favour of the wider state; second, the parameters of the other $f_0$ mesons had small corrections\cite{SBLKN-prd14}.

A further addition of decay data on $\Upsilon(mS)\to\Upsilon(nS)\pi\pi$ ($m>n,~m=2,3,4,5,~n=1,2,3$) from the ARGUS, Crystal Ball, CLEO, CUSB, DM2, Mark~II, Mark~III, BES~II, {\it BABAR}, and Belle collaborations in a combined analysis did not add any new constraints on the $f_0$ mesons, thus confirming the previous conclusions about these states\cite{SBLKN-prd14}.

On the other hand, the analysis resulted in an interesting explanation of unexpected (and even enigmatic) behavior of the dipion spectra in the decays $\Upsilon(mS)\to\Upsilon(nS)\pi\pi$ ($m>n,~m=3,4,5,~n=1,2,3$) --- a bell-shaped form in the near-$\pi\pi$-threshold region [especially for the $\Upsilon(3S)\to\Upsilon(1S) \pi^+ \pi^-$ and $\Upsilon(4S)\to\Upsilon(2S) \pi^+ \pi^-$], smooth dips near a dipion mass of 0.7~GeV in $\Upsilon(3S)\to\Upsilon(1S)(\pi^+ \pi^-,\pi^0\pi^0)$ [this implies the enigmatic two-humped shape of the dipion spectrum in these decays], of 0.6~GeV in $\Upsilon(4S,5S)\to\Upsilon(1S) \pi^+ \pi^-$ and of about 0.44~GeV in $\Upsilon(4S)\to\Upsilon(2S) \pi^+ \pi^-$, and also sharp dips of about 1~GeV in the $\Upsilon(4S,5S)\to\Upsilon(1S) \pi^+ \pi^-$ transition.
Obviously, this shape of the dipion mass distributions is explained by the interference between the $\pi\pi$-scattering and $K\overline{K}\to\pi\pi$ contributions to the final states of these decays --- by the constructive interference
in the near-$\pi\pi$-threshold region and by a destructive one in the dip regions.
Whereas the data on $\Upsilon(5S)\to\Upsilon(1S) \pi^+ \pi^-$ confirm the sharp dips near 1~GeV, the scarce data on $\Upsilon(4S)\to\Upsilon(1S) \pi^+ \pi^-$ do not allow for such a unique conclusion yet.

In turn, the consideration of the role of the individual $f_0$ resonances in making up the shape of the dipion mass distributions in these decays led to a number of interesting results. First, there are confirmed surely the sharp dips near 1~GeV in the dipion mass spectrum of the decay $\Upsilon(4S)\to\Upsilon(1S) \pi^+ \pi^-$. Furthermore, it is seen that these sharp dips in the $\Upsilon(4S,5S)$ decays are related to the $f_0(500)$ contribution in the interfering amplitudes of $\pi\pi$ scattering and the $K\overline{K}\to\pi\pi$ process.
Second, one should also note the unexpected considerable contribution of the $f_0(1370)$ to the bell-shaped form in the near-$\pi\pi$-threshold region, especially in the decays $\Upsilon(3S)\to\Upsilon(1S)(\pi^+ \pi^-,\pi^0\pi^0)$ and $\Upsilon(4S)\to\Upsilon(2S)\pi^+ \pi^-$. This is interesting because the $f_0(1370)$ is predominantly the $s{\bar s}$ state according to the earlier analysis\cite{SBLKN-prd14} and practically does not contribute to the $\pi\pi$-scattering amplitude. However, this state influences noticeably the $K\overline{K}$ scattering; e.g., it was shown that the $K\overline{K}$-scattering length is very sensitive to whether this state exists or not\cite{SKN-epja02}.
Interference of the contributions of the $\pi\pi$-scattering and analytically-continued $K\overline{K}\to\pi\pi$ amplitudes leads to the bell-shaped form of the dipion spectrum of decays $\Upsilon(mS)$ ($m=3,4,5$), which is observed in the near-$\pi\pi$-threshold region.

\section{Summary}

We performed a combined analysis of data on isoscalar $S$-wave processes $\pi\pi\to\pi\pi,K\overline{K},\eta\eta$, on the decays of the charmonia --- $J/\psi\to\phi(\pi\pi, K\overline{K})$, $\psi(2S)\to J/\psi\,\pi\pi$ --- and of the bottomonia ---
$\Upsilon(mS)\to\Upsilon(nS)\pi\pi$ ($m>n$, $m=2,3,4,5,$ $n=1,2,3$) from the ARGUS, Crystal Ball, CLEO, CUSB, DM2, Mark~II, Mark~III, BES~II, {\it BABAR}, and Belle collaborations.

It was shown that the dipion mass distributions in the two-pion transitions both of charmonia and bottomonia are explained by a unified mechanism related to contributions of the $\pi\pi$ and $K\overline{K}$ coupled channels and their interference. The role of the individual $f_0$ resonances in making up the shape of the dipion mass distributions in these decays was considered. It is shown that in the final states of these decays (except $\pi\pi$ scattering) the contribution of coupled processes, e.g., $K\overline{K}\to\pi\pi$, is important even if these processes are energetically forbidden. This is in accordance with our previous conclusions on the wide resonances\cite{SBLKN-prd14}: If a wide resonance cannot decay into a channel which opens above its mass but the resonance is strongly connected with this channel (e.g. the $f_0(500)$ and the $K\overline{K}$ channel), one should consider this resonance as a multi-channel state with allowing for the indicated channel and performing the combined analysis of the considered and coupled channels.

When describing the bottomonia decays, we did not change the resonance parameters in comparison with the ones obtained in the combined analysis of the processes $\pi\pi\to\pi\pi,K\overline{K},\eta\eta$, and charmonia decays\cite{SBLKN-prd14}.
Thus, the results of the analysis confirmed all of our earlier conclusions on the scalar mesons\cite{SBLKN-prd14}.

\section*{Acknowledgments}

This work was supported in part by the Heisenberg-Landau Program, the Votruba-Blokhintsev Program for Cooperation of Czech Republic
with JINR, the Grant Agency of the Czech Republic (Grant No. P203/15/04301), the Grant Program of Plenipotentiary of Slovak Republic at JINR, the Bogoliubov-Infeld Program for Cooperation of Poland with JINR, the Tomsk State University Competitiveness Improvement Program, the Russian Federation program ``Nauka'' (Contract No. 0.1526.2015, 3854), Slovak Grant Agency VEGA under Contract No. 2/0197/14, and by the Polish National Science Center (NCN) Grant No. DEC-2013/09/B/ST2/04382.


\end{document}